\documentstyle[12pt]{article}

\def\labelmark{}
\def\void{}
\newenvironment{formula}[1]{\def\labelname{#1}
\ifx\void\labelname\def\junk{\begin{displaymath}}
\else\def\junk{\begin{equation}\label{\labelname}}\fi\junk}%
{\ifx\void\labelname\def\junk{\end{displaymath}}
\else\def\junk{\end{equation}}\fi\junk\labelmark\def\labelname{}}
{\ifx\void\labelname\def\junk{\end{array}\end{displaymath}}
\else\def\junk{\end{array}\right.\end{equation}}
\fi\junk\labelmark\def\labelname{}\def\junk{}
}
\newenvironment{formulae}[1]{\def\labelname{#1}
\ifx\void\labelname\def\junk{\begin{displaymath}}
\else\def\junk{\begin{eqnarray}\label{\labelname}}\fi\junk}%
{\ifx\void\labelname\def\junk{\end{displaymath}}
\else\def\junk{\end{eqnarray}}\fi\junk\labelmark\def\labelname{}}

\newcommand{\beq}{\begin{formula}}
\newcommand{\eeq}{\end{formula}}
\newcommand{\beqa}{\begin{formulae}}
\newcommand{\eeqa}{\end{formulae}}
\def\BC{\bb C}
\def\_\BC{\bbi C}

\newcommand{\la}{\lambda}

\newcommand{\ep}{\epsilon}
\newcommand{\be}{\beta}
\newcommand{\ga}{\gamma}

\newcommand{\al}{\alpha}
\newcommand{\ap}{\approx}
\newcommand{\de}{\delta}
\newcommand{\lga}{\longrightarrow}

\newcommand{\Th}{\Theta}
\newcommand{\ze}{\zeta}

\begin{document}
\begin{titlepage}
\setcounter{page}{1}
\renewcommand{\thefootnote}{\fnsymbol{footnote}}

\begin{flushright}
IC/2000/161\\
hep-th/0010213
\end{flushright}

\vspace{13mm}
\begin{center}
{\Large D-dimensional Ideal Quantum Gases in $Ar^{n}+Br^{-n}$ Potential} 
\vspace{17mm}

{\large Ahmed Jellal$^{a}$ 
\footnote{E-mail: jellal@ictp.trieste.it -- jellal@youpy.co.uk }}
\,{and}\,
{\large Mohammed Daoud$^{a,b}$ 
\footnote{$^{b}$ Permanent address}}\\
\vspace{5mm}
{\em $^{a}$ High Energy Physics Section\\
 the Abdus Salam International Centre for Theoretical Physics\\
 Strada Costiera 11, 34100 Trieste, Italy} \\
\vspace{5mm}
{\em $^{b}$ Department of Physics, Faculty of Sciences, \\
 University Ibn Zohr, P.O.B. 28/S, Agadir, Morocco}
\end{center}

\vspace{5mm}

\begin{abstract}
The paper is concerned with thermostatistics of both $D$-dimensional Bose 
and Fermi ideal gases in a confining potential of type $Ar^{n}+Br^{-n}$.
The investigation is performed in the framework of the semiclassical 
approximation. Some physical quantities for such systems are derived, like 
density of states, density profiles and number of particles. Bose-Einstein 
condensation (BEC) is discussed in the high and low temperature regimes.
\end{abstract}

\vspace{8mm}
\vfill
\begin{flushleft}
\end{flushleft}

\end{titlepage}
\newpage
\setcounter{footnote}{0}
\renewcommand{\thefootnote}{\arabic{footnote}}
\renewcommand{\theequation}{\arabic{equation}}
\section{Introduction}\label{sec1}
\setcounter{section}{1}
\setcounter{equation}{0}
\indent
\indent                                                            %
In the recent past there has been increasing emphasis in quantum 
statistics of the ideal quantum gases in an arbitrary quantum 
potential since the recent experiment observations. The latter 
concerns the Bose-Einstein condensation {\cite{1}} and the Fermi 
quantum degeneracy {\cite{2}} for dilute alkali-metal atoms 
in magnetic or magneto-optical traps at very low temperature.
These results have received renewed attention and much of this 
attention is paid to the study of ideal quantum gases confining 
in different external potentials. In particular, the Bose-Einstein 
condensation. This was examined in many occasions in the presence of 
typical external potentials, for example; harmonic potential $[3-8]$, 
toroidal potential {\cite{9}}, double-well potential {\cite{10}} and in 
the presence of an impurity  {\cite{11}}. More recently, an interesting 
results have been obtained by Salasnich. Ideal quantum gases in 
$D$-dimensional space and confining in power-law potential {\cite{12}}
have been studied. Among the results derived in this paper, we note the 
condition of the Bose-Einstein condensation (BEC). Indeed, it is shown 
that BEC can set up if and only if ${D\over 2}+{D\over n}>1$. 

In this paper we wish to extend the Salasnich analysis {\cite{12}}
to cover more general results. For this task, we investigate
the ideal quantum gases trapped in the potential
\beq{.U}
U({\bf r})=Ar^{n}+Br^{-n}
\eeq
with $A, B$ strictly positives and $n$ is a non vanishing integer. 
For this type of potential, we give the thermostatistics properties 
which generalize the Salasnich ones. In the limit $B\lga 0$, we 
recover the Salasnich results. We prove that BEC occurs in the 
high temperature regime. 

This paper is organised as follows. In section $2$, we give 
the quantum distribution functions in the framework of the semiclassical
approximation for both bosons and fermions confining in a generic 
potential. Section $3$ is devoted to the generalization of the Salasnich
analysis when the particles, bosons and fermions, are trapped in the 
potential given by equation $(1)$. We will see that in the limit 
$B\lga 0$, one can recover the results obtained by Salasnich in {\cite{12}}. 
Bosons and fermions are studied separately. For the fermionic gas evolving 
in the $U({\bf r})$ potential, we give the corresponding Fermi functions, 
the finite and zero temperature momentum distribution, and the number of 
particles. In a similar way, we define the Bose function corresponding 
to particles obeying the Bose-Einstein statistics, we compute the finite 
temperature non condensed momentum distribution and finally we give the 
relation satisfied by the Bose transition temperature below which the BEC 
takes place. Since, this relation can't be solved in the 
general case, we restrict the BEC analysis in the high and low 
temperature regimes. The last section is devoted to the conclusions 
and perspectives of the present work.

\section{Prelimenaries}\label{sec2}
We shall briefly review the basic tools which will be useful to 
investigate the thermal properties of both Bose and Fermi ideal gases in a
generic confining external potential.

\subsection{Quantum distribution functions}\label{subsec1}
We start with a confined quantum gases (identical bosons or fermions).
The average number of occupation for the
$k$-mode is {\cite{13}} 
\beq{.N}
N_k={1\over e^{\be(\ep_k-\mu)}+1},
\eeq
for the fermions and 
\beq{.N}
N_k={1\over e^{\be(\ep_k-\mu)}-1},
\eeq
for bosons. In equations $(2)$ and $(3)$, $\mu$ is the chemical potential,
$\ep_{k}$ are the energy of particles, and $\be={1\over k_{B}T}$
the reciprocal temperature. Then, the total average occupation number $N$ of 
the particles is given by
\beq{.N}
N=\sum_{k}N_k, 
\eeq
which leads to fix the parameter $\mu$. For more details about the physical
meaning of the chemical potential $\mu$ see ref.{\cite{12}}. 
 
In conclusion, equations $(2)$ and $(3)$ will be written in the 
semiclassical approximation in the next subsection and will be 
constituted a starting point of our study for the bosons and 
fermions, respectively. 

\subsection{Semiclassical approximation}\label{subsec2}
In the semiclassical approximation (the thermodynamical limit), the energy 
spectrum may be considered as a continuum. This situation occurs when the 
number of particles is large and the energy level spacing is small. Thus, 
the quantum distribution functions can be replaced by the so-called 
phase-space distribution  
{\cite{12}}
\beq{.n}
n({\bf r},{\bf p})={1\over e^{\be(\ep({\bf r},{\bf p})-\mu)}\pm 1},
\eeq
where ${\bf r}=(r_1,r_2,......,r_D)$ and ${\bf p}=(p_1,p_2,......,p_D)$ 
are the position and momentum of $D$-dimensional system under consideration.
In equation $(5)$, plus and minus refer to fermions and bosons
respectively. To obtain the total distribution,
we compute the integral over all $2D$-dimensional phase-space. Then, we get 
\beq{.N}
N=\int {d^D{\bf r}d^D{\bf p}\over (2\pi\hbar)^D}\; n({\bf r},{\bf p}),
\eeq
which can be written as follows 
\beqa{.N}
N=\int d^D{\bf r}\; n({\bf r})=\int d^D{\bf p}\; n({\bf p}),
\eeqa
where the spacial and momentum distributions are given, respectively, by
\beq{.n}
n({\bf r})=\int {d^D{\bf p}\over (2\pi\hbar)^D}\; n({\bf r},{\bf p}),
\eeq
and
\beq{.n}
n({\bf p})=\int {d^D{\bf r}\over (2\pi\hbar)^D} \; n({\bf r},{\bf p}).
\eeq
Introducing the density of states
\beq{.1}
\rho(\ep)=\int {d^D{\bf r}d^D{\bf p}\over (2\pi\hbar)^D} \;
\de(\ep-\ep({\bf r},{\bf p})),
\eeq
with $\de(x)$ is the usual Dirac function. The average total occupation 
number of particles can be written as 
\beq{.N}
N=\int_0^{\infty} d\ep \;\rho(\ep)\; {1\over e^{\be(\ep-\mu)}\pm 1}.
\eeq
The phase-distribution, in the case of fermions at zero temperature 
(i.e where the chemical potential $\mu$ coincides with Fermi
energy $E_F$), takes the simplest form
\beq{.n}
n({\bf r},{\bf p})=\Th(E_F-\ep({\bf r},{\bf p})),
\eeq
where $\Th(x)$ is the well-known Heaviside function. The case of 
bosons was discussed in great detail in {\cite{12}} (see also the 
references therein). When, the external potential is taken in to 
account, the classical single-particle energy is defined as 
\beq{.2}
\ep({\bf r},{\bf p})={{\bf p}^2\over 2m}+U({\bf r}),
\eeq
from which one can prove that the semi-classical density function 
is given by 
\beq{.1}
\rho(\ep)=({m\over 2\pi\hbar^2})^{D\over 2}\; {1\over\Gamma({D\over 2})}
\int d^D{\bf r}\; (\ep-U({\bf r}))^{D-2\over 2},
\eeq
where $\Gamma(n)$ is the Gamma function. It is interesting to note 
that equation $(14)$ is valid for all external potentials. We now
close this section related to the main tools which will be used in the 
following sections investigating the thermostatistics of bosons and
fermions trapped in the potential given by eq.$(1)$.

\section{Ideal gases in the $Ar^n+Br^{-n}$ potential }
In many experiments with alkali-metals atoms, the external trap can be 
modelled by a harmonic potential {\cite{14}}. The effects of adiabatic 
change in the trap can be represented by power-law potentials {\cite{12}}.
In general, to find the momentum distribution, the Fermi temperature and 
the Bose temperature, it is necessary to specify the external potential.

\subsection{Fermionic gas}\label{subsec1}
To discuss the statistical properties of a fermionic gas embedded in 
the potential $U({\bf r})$, let us start by defining the Fermi function. 
As we will see, this definition generalizes the one given in {\cite{12}}
and allows the computation of physical quantities, of interest in the 
study of the system under consideration, as, for instance, the density of
states and number of particles.
 
We define the Fermi function as follows 
\beq{.f}
f_n^{(B)}(z)={1\over\Gamma(n)}\; \int_{0}^{\infty} dy\; y^{n-1}\;
{ze^{-ay-by^{-1}}\over 1+ze^{-ay-by^{-1}}}.
\eeq
where $a$ and $b$ are two arbitrary positive constants.
In the particular case $b=0$, we get after rescaling the variable
\beq{.f}
f_n(z)={1\over\Gamma(n)}\; \int_{0}^{\infty} dy\; y^{n-1}\;
{ze^{-y}\over 1+ze^{-y}},
\eeq
where $f_n(z)$ is the Fermi function definition {\cite{12}} for a 
fermionic system in the $U({\bf r})=Ar^{n}$ potential. We note that,
the Fermi function can be expanded, for $|z|<1$, as
\beq{.f}
f_n^{(B)}(z)={2({b\over a})^{n\over 2}\over\Gamma(n)}\;
\sum_{j=0}^{\infty}\; (-1)^{j+1}\; z^{j}\; K_n(2\sqrt{ab}),
\eeq
in terms of the modified Bessel function  $K_n(2\sqrt{ab})$ {\cite{15}}
\beq{.K}
K_n(2\sqrt{ab})= {1\over 2}({a\over b})^{n\over 2}\;
\int_{0}^{\infty} dy\; y^{n-1}e^{-ay-by^{-1}}.
\eeq

The density of states of a fermionic quantum gas in the potential 
eq.$(1)$ can be calculated from equation $(14)$. So, we have to compute 
the integral of type   
\beq{.1}
\rho^{B}(\ep)= ({m\over 2\pi\hbar})^{D\over 2}\; {1\over\Gamma({D\over 2})}
\int d^D{\bf r}\; (\ep-Ar^n-Br^{-n})^{D-2\over 2}.
\eeq
In this order, we put $y=r^n$ and solve the following equation
\beq{.y}
\ep-Ay-By^{-1}=0.
\eeq
The latter equation has two positive solutions 
\beq{.y}
u={\ep\over 2A}+\sqrt{({\ep\over 2A})^2-{B\over A}},\qquad
\al={\ep\over 2A}-\sqrt{({\ep\over 2A})^2-{B\over A}},
\eeq
when $\ep\ge 2\sqrt{AB}$ is satisfied since the radius of the sphere $r$ 
should be a real positive. Using the relation $d^D{\bf r}={D\pi^{D\over 2}
\over\Gamma({D\over 2}+1)}$ giving the $D$-dimensional unit sphere and 
the equations $(21)$, the integral $(19)$ can be written as
\beq{.1}
\rho^{B}(\ep)= ({m\over 2\pi\hbar})^{D\over 2}\;
{D\pi^{D\over 2}A^{D-2\over 2}\over n\Gamma({D\over 2})\Gamma({D\over 2}+1)}\;
\int_{\al}^{u} dy \; y^{{D\over n}-{D\over 2}} (\al+y)^{D-2\over 2}
(u-y)^{D-2\over 2},
\eeq
which can be solved using the integral representations of the 
hypergeometric functions. Finally, we obtain 
\beq{.1}
\rho^{B}(\ep)= ({m\over 2{\sqrt\pi}\hbar})^{D\over 2}\;
{D\over n}\; A^{D-2\over 2}\; (\al)^{D-2\over 2}\;
{\Gamma({D\over n}-{D\over 2}+1)\over [\Gamma({D\over 2}+1)]^2}\;
{}_2F_1(1-{D\over 2},{D\over n}-{D\over 2}+1;{D\over n};-{u\over\al}).
\eeq
We recall that the hypergeometric function ${}_2F_1(\al,\be;\ga;z)$  
is defined by {\cite{15}}
\beq{.F}
{}_2F_1(\al,\be,\ga,z)={1\over B(\be,\ga - \be)}\;
\int_{0}^{1} dy \; y^{\be -1} (1-y)^{\ga -\be -1}(1-zy)^{-\al},
\eeq
where $Re\ga>0$, $Re\be>0$ and 
\beq{.B}
B(\al,\be)={\Gamma(\al)\Gamma(\be)\over\Gamma(\al+\be)}=B(\be,\al),
\eeq
is the so-called beta function or Euler's integral of the first kind, 
which admits the following integral representation
\beq{.B}
B(\al,\be)=\int_{0}^{1} dt\; t^{\al -1} (1-t)^{\be -1}.
\eeq
It is interesting to remark that the following relation is satisfied 
\beq{.F}
\lim_{B \to 0}\; (\al)^{\la}\;
{}_2F_1(1-{D\over 2},{D\over n}-{D\over 2}+1;{D\over n};-{u\over\al})\;=
\; ({\ep\over A})^{{D\over 2}-1}\;
{\Gamma({D\over 2})\Gamma({D\over n}+1)\Gamma({D\over n})\over
\Gamma({D\over 2}-1)\Gamma({D\over n}-{D\over 2}+1)
\Gamma({D\over n}+2)}.
\eeq
Using the latter relation, the density of states, in the limit $B\lga 0$,
is given by 
\beq{.1}
\rho(\ep)= ({m\over 2\hbar^2})^{D\over 2}\; ({1\over A})^{D\over n}\;
{\Gamma({D\over n}+1)\over\Gamma({D\over 2}+1)\Gamma({D\over n}+{D\over 2})}
\;\ep^{{D\over n}+{D\over 2}+1}
\eeq
which coincides with the result obtained in reference {\cite{12}}.

Using the above tools, other interesting properties of the present 
system (fermions) may be determined in a simple way: 

{\underline {\bf i- The finite temperature momentum distribution}} 
\beq{.n}
n({\bf p})={1\over (2{\sqrt\pi}\hbar)^{D}}\;
{\Gamma({D\over n}+1)\over\Gamma({D\over 2}+1)}\;
f_{D\over n}^{(B)}(e^{\be(\mu-{{\bf p}^2\over 2m})}).
\eeq
{\underline{\bf ii- The zero temperature momentum distribution}}
\beq{.n}
n({\bf p})={1\over (2{\sqrt\pi}\hbar)^{D}}\;
{1\over\Gamma({D\over 2}+1)}\;\Theta(E_F-{{\bf p}^2\over 2m})\;
[{E_F-{{\bf p}^2\over 2m}\over 2A}+[({E_F-{{\bf p}^2\over 2m}
\over 2A})^2-{B\over A}]^{1\over 2}]^{D\over n},
\eeq
where we request that $p\leq [2m(\ep-2\sqrt{AB})]^{1\over 2}$ for 
the right term makes sense.\\
{\underline{\bf iii- Number of particles}}\\
The number of particles is given by the following integral 
\beq{.N}
N={1\over\Gamma({D\over 2}+1)}\;\int {d^{D}{\bf p}\over(2\pi\hbar)^{D}}\;
\Theta(E_F-{{\bf p}^2\over 2m})\; [{E_F-{{\bf p}^2\over 2m}\over 2A}+
[({E_F-{{\bf p}^2\over 2m}\over 2A})^2-{B\over A}]^{1\over 2}]^{D\over n},
\eeq
which can be reorganized as   
\beq{.N}
N={1\over (2{\sqrt\pi}\hbar)^{D}}\; {D\pi^{D\over 2}\over
[\Gamma({D\over 2}+1)]^2}\; \int_{0}^{[2m(\ep-2\sqrt{AB})]^{1\over 2}} 
dp\; p^{D-1}\; [{E_F-{{\bf p}^2\over 2m}\over 2A}+
[({E_F-{{\bf p}^2\over 2m}\over 2A})^2-{B\over A}]^{1\over 2}]^{D\over n}.
\eeq
A calculation more or less complicated, gives the following expression 
for the number of particles 
\beq{.N}
N=M\; f(\ga)\; B({D\over 2},{D\over 2})\; [f_1(\ga)\; {}_2F_1(+1)-
f_1^{-1}(\ga)\; {}_2F_1(-1)],
\eeq
where
$$M={D({m\over 2})^{1\over 2}\over\hbar^{2}}\; {(AB)^{{D\over 2n}+
{D\over 4}}\over A^{D/n}},$$
$$f(\ga)=(\ga^{2}-1)^{D-1\over 2}\;[\ga-(\ga^{2}-1)^{1\over 2}]^{{D\over n}-
{D\over 2}},$$
$$f_1(\ga)=\ga-(\ga^{2}-1)^{1\over 2},$$ 
$$f_1^{-1}(\ga)={1\over f_1(\ga)},$$
$$\ga={E_F-{{\bf p}^2\over 2m}\over 2A},$$
$${}_2F_1(\pm1)={}_2F_1(-{D\over n}+{D\over 2}\mp1;{D\over 2};D,-
{2(\ga^{2}-1)^{1\over 2}\over \ga-(\ga^{2}-1)^{1\over 2}}),$$
and $B({D\over 2},{D\over 2})$ is given by eqs.(25,26). 

The number of particles is expressed in terms of the hypergeometric 
and beta functions. It is clear that this relation is complicated and we
believe that a more detailed study of it can generate more interesting 
results. We will return to this matter in a subsequent paper.

To close this subsection, we note that the results derived in ref.{\cite{12}}
related to the fermionic gas, can be reproduced just by taking the limit
$B\lga 0$.

\subsection{Bosonic gas}\label{subsec2}
We now examine the condensation of a system of bosons confining 
in the potential $U({\bf r})$ (eq.$(1)$). We start by defining the Bose 
function as follows 
\beq{.f}
g_n^{(B)}(z)={1\over\Gamma(n)}\;\int_{0}^{\infty} dy\; y^{n-1}\;
{ze^{-ay-by^{-1}}\over 1-ze^{-ay-by^{-1}}}.
\eeq
For $|z|<1$, we have
\beq{.f}
g_n^{(B)}(z)={2({b\over a})^{n\over 2}\over\Gamma(n)}\;
\sum_{j=1}^{\infty}\;z^{j}\; K_n(2j\sqrt{ab}),
\eeq
where $K_n(2\sqrt{ab})$ is given by eq.$(18)$.
For $B\lga 0$, we show that 
\beq{.f}
g_n(z)={1\over\Gamma(n)}\;\int_{0}^{\infty} dy\; y^{n-1}\;
{ze^{-y}\over 1+ze^{-y}}, 
\eeq 
which is nothing but the definition of the Bose function given in 
{\cite{12}} corresponding to the bosons in the potential
$U({\bf r})=Ar^n$.

Using the definition given by eq.$(34)$, one obtain the following results: \\
{\underline{\bf i- The finite temperature non condensed momentum 
distribution}} 
\beq{.n}
n({\bf p})={1\over (2{\sqrt\pi}\hbar)^{D}}\;
{\Gamma({D\over n}+1)\over\Gamma({D\over 2}+1)}\;
g_{D\over n}^{(B)}(e^{\be(\mu-{{\bf p}^2\over 2m})}).
\eeq
\underline{\bf ii- The Bose transition temperature} \\
To obtain the Bose-temperature, we have to calculate the number 
of particles 
\beq{.N}
N=\int d^{D}{\bf p}\; n({\bf p}).
\eeq
As in the case of ordinary bosons (without interaction), we define the 
Bose-temperature by taking the chemical potential $\mu=0$. By 
introducing $\mu=0$ in eq.$(37)$, we get 
\beq{.N}
N= {D\pi^{1\over 2}\over (2{\sqrt\pi}\hbar)^{D}}\;
{\Gamma({D\over n}+1)\over[\Gamma({D\over 2}+1)]^2}\;
\int_{0}^{\infty}dp\; p^{D-1}\; g_{D\over n}^{(B)}
(e^{-\be{\bf p}^2\over 2m}).
\eeq
With the help of eq.$(35)$, it is easy to show that 
\beq{.N}
N= 2({B\over A})^{D\over 2n}\; ({2m\over\be})^{D\over 2}\;
{D\pi^{1\over 2}\over (2{\sqrt\pi}\hbar)^{D}}\;
{\Gamma({D\over n}+1)\over\Gamma({D\over 2}+1)\Gamma({D\over 2}+1)}\;
\sum_{j=1}^{\infty}\; {K_{D\over n}(2j\be\sqrt{AB})\over j^{D\over 2}}.
\eeq
The latter equation gives a relation between the number of particles
(or density) and Bose-temperature below which we obtain a Bose-Einstein
condensation. Due to the complicated sum appearing in $(40)$, the Bose 
temperature can't be calculated, in general, in a simple way. However, in
the some particular situations, the solutions of this equation can be 
found. Indeed, in the high and low temperature regimes (which are two
interesting situations from an experimental point of view) some 
information about BEC can be derived:\\
{\underline{\textsf{High temperature}}}: This case corresponds to 
$\be\lga 0$. In this limit the functions $K_{D\over n}(2j\be\sqrt{AB})$ 
take the form {\cite{16}}
\beq{.K}
K_{D\over n}(2j\be\sqrt{AB})\; \ap {1\over 2}\; \Gamma({D\over n})\;
(2j\be\sqrt{AB})^{-{D\over n}}.
\eeq
Therefore, eq.$(40)$ implies
\beq{.N}
N= ({m\over 2\hbar^{2}})^{D\over 2}\;
{\Gamma({D\over n}+1)\over\Gamma({D\over 2})}\;
{\ze({D\over 2}+{D\over n})\over\be^{{D\over 2}+{D\over n}}}.
\eeq
This result is similar to the one found in ref.{\cite{12}}. From this relation 
we conclude that just restricting ourselves to high temperature limit, we find 
the Salasnich analysis related to the BEC. Indeed, it is shown that the 
latter takes place when the condition ${D\over 2}+{D\over n}>1$ is 
satisfied. Strictly speaking, we obtain the condensed fraction as follows
\beq{.N}
{N_0\over N}= 1- ({T\over T_B})^{{D\over 2}+{D\over n}},
\eeq
where $N_0$ is the number of particles occupying the single-particle 
ground-state of the system when the temperature is below $T_B$. \\
{\underline{\textsf{Low temperature}}}: Equivalently to $\be\lga \infty$, then 
we have {\cite{15}}
\beq{.K}
K_{D\over n}(2j\be\sqrt{AB})\; \ap \; 0.
\eeq
In that case, we get a number of particles $N\lga 0$ corresponding to
a temperature $T\lga 0$. This result is compatible with the literature
{\cite{13}}.

\section{Conclusion}\label{sec3}
In this work, we concentrated on the influence of confining potential 
$U({\bf r})=Ar^{n}+Br^{-n}$ on the main statistical properties of the 
ideal quantum gases (bosons or fermions) in $D$-dimensions. We discussed 
the derivation of the density of states, spacial and momentum distributions 
in the thermodynamical limit. For fermions, we have calculated the Fermi 
energy and for bosons, the phenomenon of Bose-Einstein condensation is 
discussed in terms of the reciprocal temperature $(\be)$. Two situations 
were considerd $(\be\lga 0)$ and $(\be\lga\infty)$ corresponding to high and 
low temperature, respectively. In the limit $B\lga 0$, our results
reproduce the Salasnich ones ref.{\cite{12}} concerning the confining 
power-law potential. 

An important result of the present work concerns the Bose-Einstein 
condensation, in the high temperature domain, which occurs when 
${D\over 2}+{D\over n}>1$, where $D$ is the space dimension and $n$ 
the exponent appearing in the expression of $U({\bf r})$. Another result 
concerns to the low temperature case where the number of particles vanishes.
 
To finalize this paper, it should be noted that the present investigation
of the ideal quantum gases embedded in the potential $U({\bf r})$ 
(eq.$(1)$) in $D$-dimensions, constitutes now a very interesting topic to 
learn more about Bose-Einstein condensation. In fact, as we have indicated 
above, the BEC should be investigated in more detail in general case solving 
equation $(40)$. We believe that more important informations can be obtained 
in this case. We will return on this subject in a forthcoming work {\cite{17}}.

\section*{Acknowledgment}
A. Jellal is grateful to Prof. S. Randjbar-Daemi for the kind invitation
to visit the High Energy Section of the Abdus Salam International
Centre for Theoretical Physics (AS-ICTP). M. Daoud acknowledges hospitality 
of (AS-ICTP) and would like to thank Prof. Yu Lu. The authors are 
thankful to Prof. Salasnich for the discussions about his work 
reference {\cite{12}} and for his kind comment concerning this paper.
The authors are greatly indebted to Prof. G. Thompson for reading
the manuscript.
%

%


\begin{thebibliography}{99}

\bibitem{1}M.H. Anderson, J.R. Ensher, M.R. Matthews, C.E. Wieman and E.A. 
Cornell, {\it Science} {\bf 269} (1995) 198.
\bibitem{2}B. DeMarco and D.S. Jin, {\it Science} {\bf 285} (1999) 1703.
\bibitem{3}L. Salasnich, {\it Mod. Phys. Lett.} {\bf B11} (1997) 1249,
\textsf{quant-ph/9712030}.
\bibitem{4}L. Salasnich, {\it Mod. Phys. Lett.} {\bf B12} (1998) 649,
\textsf{quant-ph/9801001}.
\bibitem{5}E. Cerboneschi, R. Mannella, E. Arimondo and L. Salasnich, 
{\it Phys. Lett.} {\bf A249} (1998) 245, \textsf{cond-mat/9809371}.
\bibitem{6}L. Salasnich, {\it Phys. Rev.}{\bf A61} (2000) 015601, 
\textsf{cond-mat/9909070}.
\bibitem{7}L. Salasnich, {\it Int. J. Mod. Phys. Lett.} {\bf B14} (2000) 1, 
\textsf{cond-mat/9908147}.
\bibitem{8}L. Salasnich, {\it Phys. Lett.} {\bf A266} (2000) 187, 
\textsf{cond-mat/0001087}.
\bibitem{9}L. Salasnich, A. Parola and L. Reatto, {\it Phys. Rev.}
{\bf A59} (1999) 2990, \textsf{cond-mat/9901121}.
\bibitem{10}L. Salasnich, A. Parola and L. Reatto, {\it Phys. Rev.}
{\bf A60} (1999) 4171, \textsf{cond-mat/9908146} .
\bibitem{11}P. Giacconi, Fabio Maltoni and R. Soldati, {\it ``Bose-Einstein 
Condensation in the Presence of an Impurity''}, \textsf{cond-mat/0007395}.
\bibitem{12}L. Salasnich, {\it ``Ideal Quantum Gases in $D$-dimensional 
Space and Power-Law Potentials''}, to appear in {\it J. Math. Phys.},
\textsf{math-ph/0008030}.
\bibitem{13} C. Garrod, {\it ``Statistical Mechanics and Thermodynamics''}, 
(Oxford University Press, 1995).
\bibitem{14}F. Dalfovo, S. Giorgini, L.P. Pitaevskii and S. Stringari,
{\it Rev. Mod Phys.} {\bf 71} (1999) 463.
\bibitem{15}I.S. Gradshteyn and I.M. Ryzhik, {\it ``Table of Integrales, 
Series and Products''},
(Academic Press, Orlando 1980).
\bibitem{16}M. Abramowitz and I.A. Stegun, {\it ``Handbook of Mathematical 
Functions''}, (Dover, New York 1980).
\bibitem{17}A. Jellal and M. Daoud, work in preparation.
\end{thebibliography}
\end{document}